\documentclass[aip,rsi,graphicx,superscriptaddress]{revtex4-1} 
\usepackage{amsmath}
\usepackage{amssymb, bbm, graphicx,color}
\usepackage{mathtools}

\begin{document}
\title{ Direct measurement of fast transients by using boot-strapped waveform averaging}

\author{Mattias Olsson}
\affiliation{Department of Electrical and Information Technology, Lund University, Ole R\"omers v\"ag 3, 22363, Lund, Sweden}
\author{Fredrik Edman}
\affiliation{Department of Electrical and Information Technology, Lund University, Ole R\"omers v\"ag 3, 22363, Lund, Sweden}
\author{Khadga Jung Karki}
\email{Khadga.Karki@chemphys.lu.se}
\affiliation{Chemical Physics, Lund University, Naturevetarv\"agen 16, 22362, Lund, Sweden}

\begin{abstract}
An approximation to coherent sampling, also known as boot-strapped waveform averaging, is presented. The method uses digital cavities to determine the condition for coherent sampling. It can be used to increase the effective sampling rate of a repetitive signal and the signal to noise ratio simultaneously. The method is demonstrated by using it to directly measure the fluorescence lifetime from rhodamine 6G by digitizing the signal from a fast avalanche photodiode. The obtained lifetime of 4.4$\pm0.1$ ns is in agreement with the known values. 

\end{abstract}


\maketitle

\section{Introduction}
The analog to digital converters (ADCs) that are used in mordern signal digitizers have fixed sampling rates, that can range up to few giga samples per second for digitizers with 8 bit ADCs. In order to further increase the sampling rate, one uses some form of interleaving methods. The interleaving techniques may be classified into two groups, one based on hardware and other based on signal processing. 

The hardware based interleaving uses multiple ADCs, which digitize the signal in parallel. In this technique, the signal is simultaneously channeled to the different ADCs.~\cite{TI_INTERLEAVING} The ADCs are triggered with slight phase difference such that they sample different sections of the signal. Finally, the data stream from all the ADCs are merged based on the trigger sequence. In this technique, the effective sampling rate is the multiple of the sampling rate of one ADC and the number of the ADCs used in parallel.

Random interleaved sampling (RIS) is another interleaving method based on the signal processing. It is specialized feature used in some of the digital sampling oscilloscopes (DSO)~\cite{LECROY_RIS}. In RIS, the signal is sampled multiple times by changing the trigger position using a time-to-digital converter (TDC). The value from the TDC is used to arrange and interleave the different waveforms. If $n$ waveforms are interleaved then the end result is a single waveform sampled at an effective sampling rate that is $n$ times the digitization speed of the ADC. One of the drawbacks of this technique is that it requires TDC for controlling the trigger, which is not available in general ADC boards.   

Coherent sampling is yet another method that can be used to increase the effective sampling rate of repetitive signal. In this method, the sampling condition that fulfills the condition in equation~\eqref{EQ1} 
\begin{equation}\label{EQ1}
\frac{f_{IN}}{f_s} = \frac{M}{N},
\end{equation}
where $f_s$ is the sampling frequency, $f_{IN}$ is the signal frequency, $M$ is the integer number of cycles in the data record and $N$ is the integer number of samples in the record.~\cite{INTERSIL} Coherent sampling is primarily used in sinewave testing of ADCs. Nevertheless,  the $M$ cycles of the waveform obtained by coherent sampling can be unwarped to increase the effective sampling rate.  However, finding the right condition for coherent sampling is complex.~\cite{INTERSIL}

Here, we implement an approximation of coherent sampling based on the digital cavities, which we call boot-strapped waveform averaging. The technique is simple, does not require external trigger from the signal source during digitization, can be used to analyze any repetitive signals and the effective sampling rate can be varied. We demonstrate the technique by using it to measure the lifetime of fluorescence from a dye molecule, Rhodamine 6G, with a time precision of 10 ps using a digitizer with the normal sampling rate of 1 GSa/s.

\section{Algorithms of boot-strapped wave-form averaging}

The boot-strapped wave-form averaging is an approximation to the coherent sampling. In order to describe the algorithm, we assume that the signal is digitized at regular intervals, $\tau$. In the method, one first approximates the window that holds complete $M$ cycles by using using the digital cavities. The data stream is folded $m$ times into the cavities of varying length. When the cavity length approximately matches the coherent sampling condition, the signal within the cavity enhances. The folding of the data in the cavity also reduces the random noise by $\sqrt{m}$. If the length of the cavity for coherent build up of the signal is $N$, then the approximate time period of the waveform is given by 
\begin{equation}
T = \frac{N\tau}{M}.
\end{equation}
Using $T$, one can ``boot-strap" the different cycles of the waveform in the digital cavity to generate one waveform that is effectively sampled at the regular intervals of $\tau/M$.  

\section{Measurement of fluorescence lifetime using boot-strapped waveform averaging}
\begin{figure}[h]
\includegraphics[width=6in]{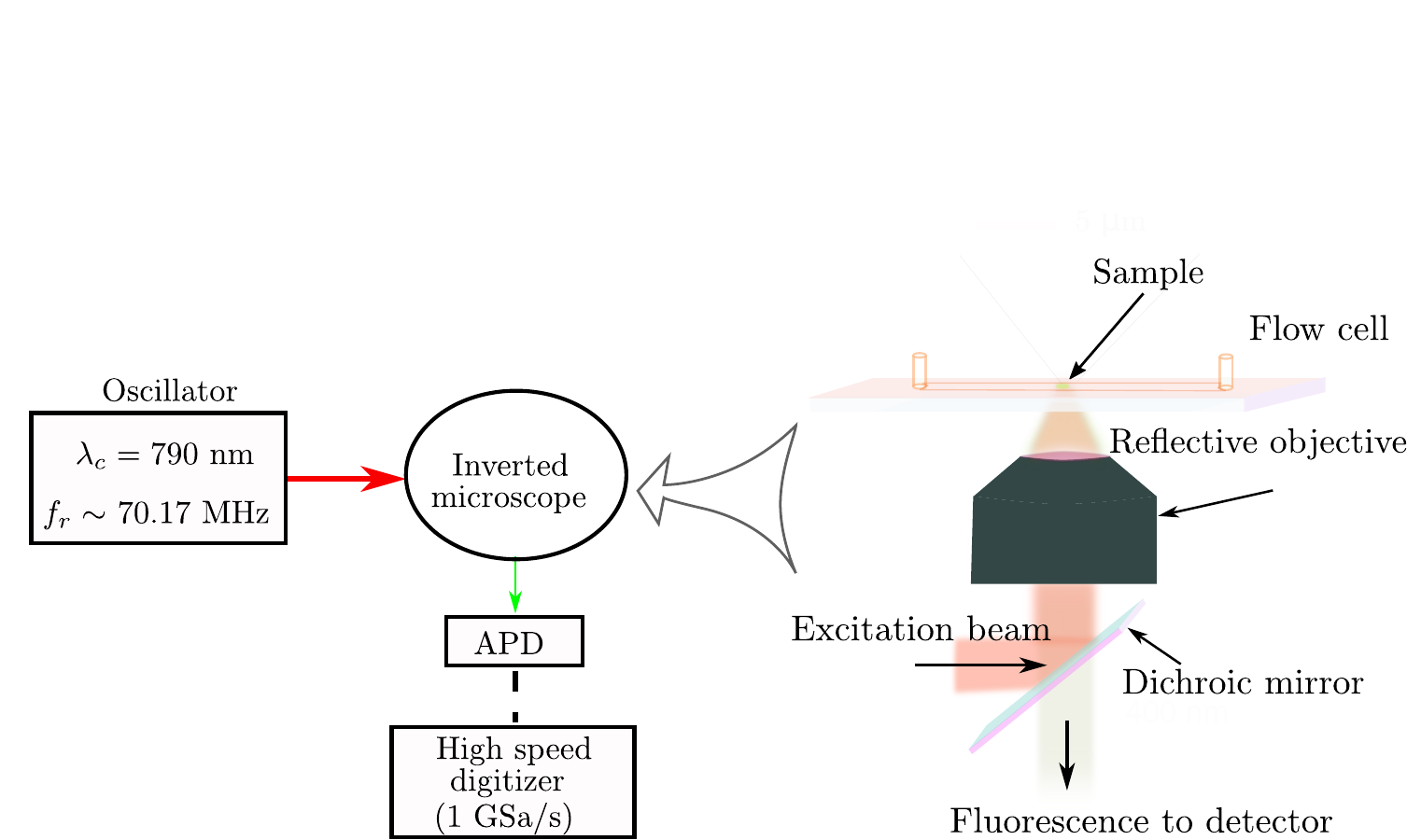}
\caption{Experimental setup. Laser source is a mode-locked oscillator. The center wavelength is 790 nm, repetition rate is about 70.17 MHz and the duration of the laser pulses is about 10 fs. The schematic of the microscope is shown in the right. Two-photon fluorescence from the sample is detected by a fast avalanche photodiode. The signal from the photodiode is digitized by a fast digitizer and saved for further signal processing.}
\label{FIG1}
\end{figure}

In order to demonstrate an implementation of the algorithm, we have used it to measure the lifetime of fluorescence from Rhodamine 6G. The schematics of the experimental setup is shown in Figure \ref{FIG1}. The optical setup has been described elsewhere.~\cite{KARKI_2016A,KARKI_2016C} A Ti:Sapphire oscillator (Synergy, Femtolasers) is used as the optical source. The center wavelength of the output is at 790 nm and the pulse duration is about 10 fs. The oscillator generates about 70.17 million pulses every second. An inverted microscope (Nikon Ti-S) is used to focus the laser beam onto the sample in a flow-cell. The microscope setup has been described elsewhere.~\cite{KARKI_2016B} A millimolar solution of rhodamine 6G in water is used as the sample. The fluorescence from the sample excited by two-photon absorption is directed to a fast avalanche photodiode (APD) (APD210, MenloSystems GmbH). The APD has a bandwidth of 1.6 GHz. The signal from the APD is digitized by a fast digitizer (ATS9870, Alazartech) at the rate of 1 GSa/s (giga samples per second).   

\begin{figure}[h]
\includegraphics[width=5in]{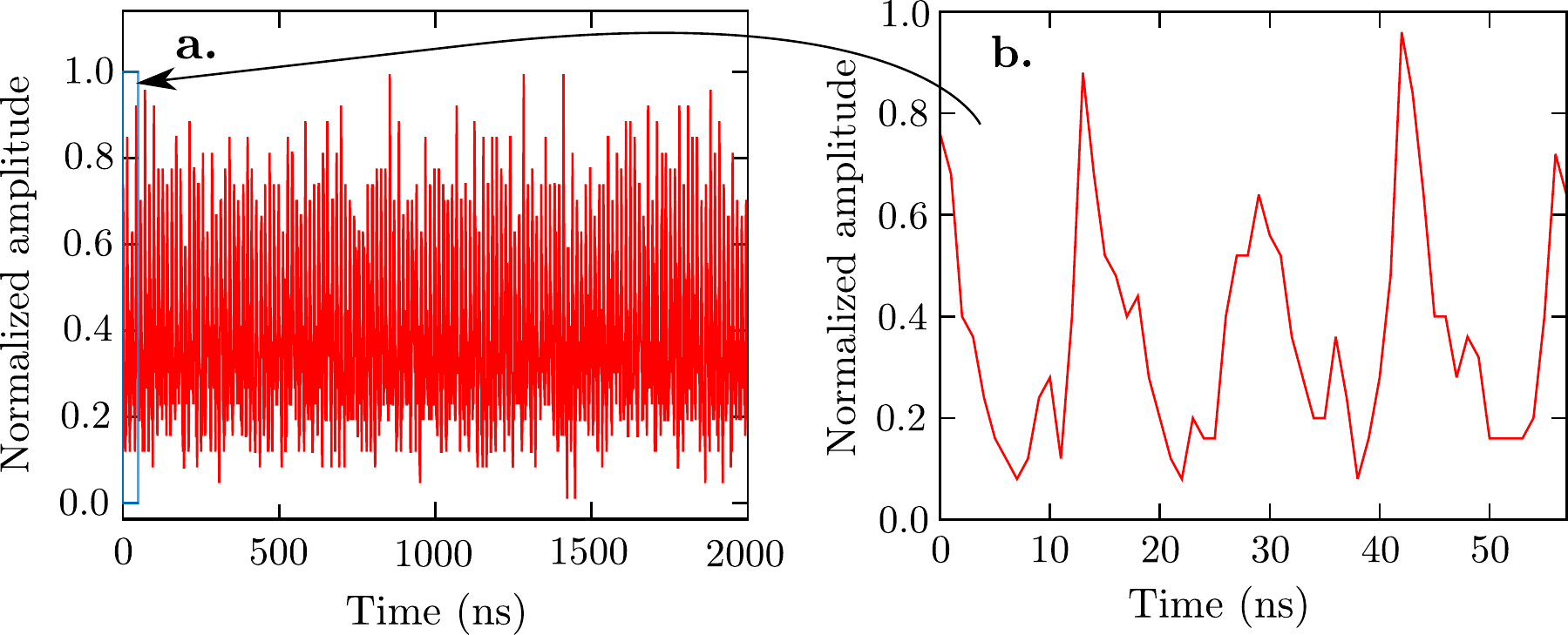}
\caption{Raw data. (a) shows 2000 samples and (b) shows first 57 samples.}
\label{FIG2}
\end{figure}

A sample of the raw digitized signal is shown in Figure \ref{FIG2} (\textbf{a.}). Figure \ref{FIG2}(\textbf{b.}) shows the first 57 data points. As shown in the figure, the data has a poor temporal resolution of 1 ns and is rather noisy because of which one does not observe the clear decay profile of the fluorescence signal. Previously, we have shown that digital cavities can be used to improve the signal to noise ratio in sinusoidal signals.~\cite{KARKI_2013A,KARKI_2013C} Here, although the signal is not sinusoidal, it is repetitive. Nevertheless, a repetitive signal can be decomposed into a fundamental sinusoidal signal and its harmonics, one can still use the algorithm of digital cavity to improve the signal to noise ratio. 

\begin{figure}[h]
\includegraphics[width=3in]{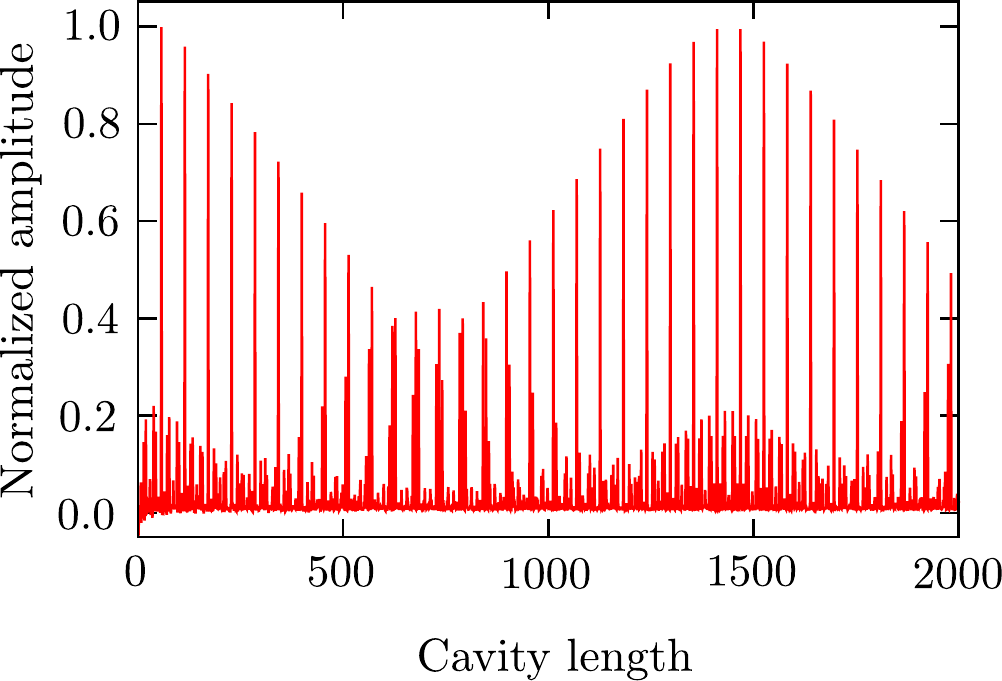}
\caption{Amplitude of the signal filtered by digital cavities of various lengths.}
\label{FIG3}
\end{figure}

A digital cavity is defined by its length, i.e. the number of bins in which the data points are accumulated. The application of the cavity refers to the repeated folding of the data stream in the cavity.~\cite{KARKI_2013A} If the period of the repetitive signal matches the cavity length it gets enhanced by the number of times the data is folded, otherwise the signal gets averaged out. In order to find the period after which the signal repeats, we vary the cavity length.  Figure~\ref{FIG3} shows the amplitude of the signal after averaging 80 times in the digital cavities of varying length. We observe maximum amplitudes when the length of the cavity is 57 ns or 1411 ns. The signal approximately fulfills the condition of coherent sampling at these lengths of the cavity. 
 
\begin{figure}[h]
\includegraphics[width=5in]{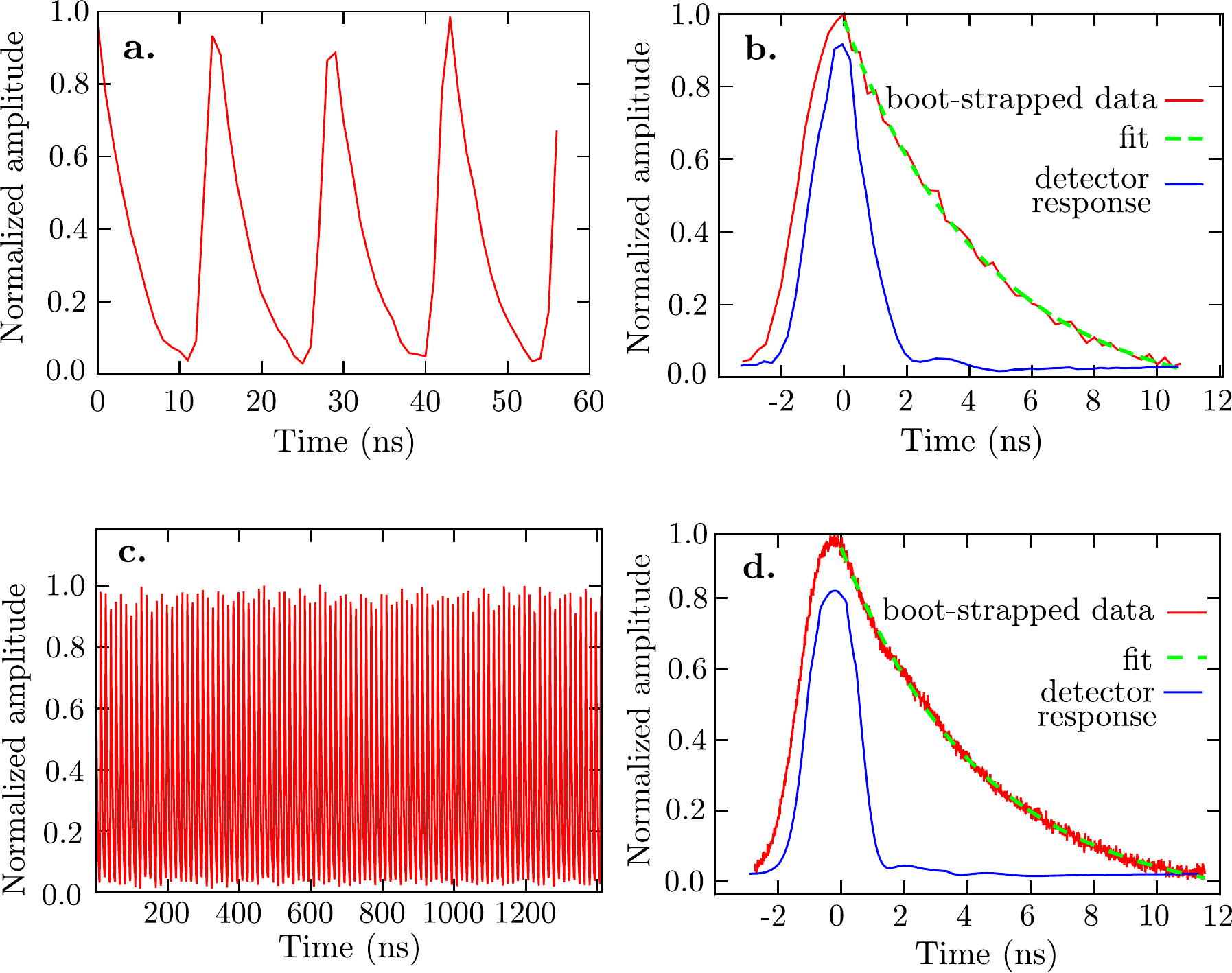}
\caption{(a) and (c) are Coherently averaged waveforms using cavities of lengths 57 and 1411 ns, respectively. (b) and (d) show bood-strapped data of (a) and (c), respectively, together with exponential fits and detector response.}
\label{FIG4}
\end{figure}

Figure~\ref{FIG4}(\textbf{a.}) shows the average signal obtained from the digital cavity of length 57 ns. Compared to the raw data in Figure~\ref{FIG2}(\textbf{b.}), the signal to noise ratio has improved considerably such that the fast decay of fluorescence is clearly visible. There are four cycles of the fluorescence transients, which are not identical. The difference in the four transients observed in figure~\ref{FIG4}(\textbf{a.}) is due to the mismatch in the repetition rate of the laser and the rate of digitization of the signal. As the repetition rate of the laser is $\sim$70.17 MHz, the transients repeat with a period of about 14.25 ns. However, as the sampling is done every ns. Thus, the data sampled at 15th nanosecond is phase shifted by -0.25 ns compared to the first sample. Note that the period of repetition can also be directly obtained from the cavity averaged signal. As four cycles of the transients have a length of 57 ns, the period of each cycle has to be 14.25 ns. Finally, we merge or boot-strap the different transients  by taking into account the relative phase shifts. Figure~\ref{FIG4}(\textbf{b.}) shows the merged transient labeled as ``boot-strapped data". The data has been shifted such that the maximum of the signal is at time zero. The effective sampling rate of the data is 4 GSa/s and the corresponding temporal resolution is 250 ps.  The dashed line is the exponential decay fit of the data, which has a decay constant of 4.5$\pm$0.2 ns. The figure also shows the response of the detector to attenuated laser beam, which too is  obtained by using boot-strapped waveform averaging. 

The effective sampling rate can be increased further by using the digital cavity of longer length. Figure~\ref{FIG4}(\textbf{c.}) shows the average signal obtained from the digital cavity of length 1141 ns. The appropriate length of the cavity is chosen based on the second maxima amplitude in figure~\ref{FIG2}. The averaged signal in the cavity has 99 cycles, from which we obtain the repetition period of $\sim$14.2525 ns. Here, the subsequent transients have a phase shift of -0.2525 ns, based on which we merge all the transients. The merged transient in figure~\ref{FIG4}(\textbf{d.}) has the effective sampling rate of 99 GSa/s and the corresponding temporal resolution of about 10 ps. The time constant of the exponential decay (also known as the lifetime of the fluorescence) is 4.4$\pm 0.1$ ns. The lifetimes obtained from figures~\ref{FIG4}(\textbf{b.} and \textbf{d.}) agree with the known lifetime of rhodamine 6G in water (about 4.08 ns).~\cite{FL_RHODAMINE} Thus, in this case the cavity of only 57 bins, which gives effective sampling rate of 4 GSa/s, is good enough to evaluate the fluorescence lifetime. However, for faster transients one would require longer cavity and correspondingly higher effective sampling rate. 

There are many methods, such as time correlated single photon counting (TCSPC)~\cite{CONNOR_1984}, phase modulation~\cite{RIMAI_1977,LIMKEMAN_1983}, gated fluorescence detection~\cite{BRODY_1957,BROWN_1963} and direct recording of full transients using fast digitizers~\cite{MYCEK_2001,NISHIOKA_2003,MARCU_2004}, that can used to measure the lifetime of fluorescence. TCSPC is the most commonly used technique, which has  comparatively advantageous features such as low systematic errors and best signal-to-noise ratio.~\cite{CONNOR_1984} However, the measurement time of a single transient is rather long, ~\cite{FRENCH_2007,DUNSBY_2012,DUNSBY_2014} which has been a disadvantage in applications that require fast measurements.~\cite{MARCU_2012} 

More recently, direct recording of the fluorescence transients using fast detectors and data acquisition systems have been used as an alternative lifetime measurement technique in clinical applications.~\cite{MYCEK_2001,NISHIOKA_2003,MARCU_2004,MARCU_2012}  Direct recording of the transients is particularly advantageous in samples that are highly fluorescent. In this case, it is even possible to record a transient in real time. Previously, sampling oscilloscopes have been used to record the data at the sampling rate of about 5 GSa/s (giga samples per second), which gives a temporal resolution in data acquisition of about 200 ps.~\cite{MYCEK_2001,NISHIOKA_2003,MARCU_2004,MARCU_2012} Traditional interleaved sampling methods have been used to improve the effective temporal resolution.~\cite{GILL_2007} However, such methods are time consuming,~\cite{GILL_2007} which offsets the benefits of fast measurement. The technique we have presented in this note provides a new method to rapidly carry out wave-form averaging and interleaving without slowing the measurements.  In our measurements, the data used to carry out waveform averaging with the cavities of lengths 57 and 1141 have data acquisition times of about 4.6 $\mu$s and 113 $\mu$s, respectively.   

Although, we have shown the use of boot-strapped waveform averaging for the measurement of fluorescence lifetimes, the method is general and can be used to increase the effective sampling rate in any repetitive signal.

\section{Conclusion}
We have presented an approximate method of coherent sampling and waveform averaging to increase the signal to noise ratio in a repetitive signal and at the same time increase the effective sampling rate. We have demonstrated its use by measuring the fluorescence lifetime of rhodamine 6G. The method is general and can be used to rapidly analyze any repetitive signal.

\textbf{Acknowledgments}

Financial support from Lund University Innovation System and VINNOVA is gratefully acknowledged. 

%




\end{document}